\documentstyle[emulateapj,psfig]{article}

\makeatletter

\newenvironment{inlinefigure}{%
\def\@captype{figure}%
\noindent\begin{minipage}{0.999\linewidth}\begin{center}}
{\end{center}\end{minipage}\smallskip}
\makeatother

\lefthead{Songaila}

\begin{document}

\submitted{ApJ, 561, L153 : revised 3/2/02}
\title{The Minimum Universal Metal Density Between Redshifts 1.5 and 5.5}
\author{Antoinette Songaila\altaffilmark{1}} 
\affil{Institute for Astronomy, University of Hawaii, 2680 Woodlawn Drive,
  Honolulu, HI 96822\\}

\altaffiltext{1}{Visiting astronomer, W. M. Keck
  Observatory, jointly operated by the California Institute of Technology and
  the University of California.}

\slugcomment{ApJ, 561, L153 (2001); revised 3/2/02}

\begin{abstract}
It appears that the Lyman $\alpha$\ forest is becoming thick at a redshift of
about 5.5, cutting off the higher redshift intergalactic medium from view in
neutral hydrogen.  However, the effects of star formation at higher redshift
are still readable in the intergalactic metal lines.  In this paper I use
observations of 32 quasars with emission redshifts in the range 2.31 to 5.86
to study the evolution of the intergalactic metal density from $z = 1.5$\ to
$z = 5.5$.  The \ion{C}{4} column density distribution function is consistent
with being invariant throughout this redshift range.  From direct integration,
I determine $\Omega_{\rm CIV}$\ to be in the range $(2.5 - 7) \times
10^{-8}$\ and $\Omega_{\rm SiIV}$\ in the range $(0.9 - 3) \times 10^{-8}$
between $z = 1.5$\ and $z = 5$.  The metallicity at $z = 5$\ exceeds $3.5
\times 10^{-4}$, which in turn implies that this fraction of the universal
massive star formation took place beyond this redshift.  This is sufficient to
have ionized the intergalactic medium.

\end{abstract}


\section{Introduction} \label{intro}

The epoch of reionization is the ultimate depth to which we can directly
investigate neutral hydrogen in the intergalactic medium (IGM).  Whether or
not it signals the onset of reionization, the increasing optical depth of the
Lyman $\alpha$\ forest at redshifts above 5 (Becker et al.\ 2001; Djorgowski
et al.\ 2001; Fan et al.\ 2000b, 2001c; Songaila et al.\ 1999) severely limits
our ability to study the IGM with neutral hydrogen at these times.  However,
we can continue to study the metal lines in the quasar spectra, and in
particular the \ion{C}{4} `forest'.  These absorption lines can be used to
infer the history of early star formation by mapping the evolution of the
universal density of metals with redshift.  The contribution to the universal
metallicity arising from the diffuse IGM has been extensively analyzed at
redshift $z \sim 3$\ (Cowie et al. 1995, Tytler et al. 1995, Songaila 1997,
Songaila 1998, Ellison et al.  2000) and some information is also available
from investigations of \ion{O}{6} at lower redshift (Burles \& Tytler 1996,
Tripp, Savage \& Jenkins 2000).  However, to date there has been no systematic
analysis of the variation of the intergalactic metal density as a function of
redshift.

Most interpretations of the origin of the $z = 3$\ metallicity have focussed
on ejection or stripping of material from small galaxies that formed at higher
redshifts ($z > 5$) (Gnedin \& Ostriker 1997, ; Gnedin 1998; Madau, Ferrara \&
Rees 2001).  It is also possible that the metals may originate in genuine
Population III star formation at very high redshift (e.g., Carr, Bond \&
Arnett 1984; Ostriker \& Gnedin 1996; Haiman \& Loeb 1997; Abel et
al. 1998). In these scenarios most of the IGM metals would be in place at the
highest currently observable redshifts. More recently, Steidel has suggested
that the observed metals may instead originate in extremely fast winds from
Lyman break galaxies at $z = 3$\ (e.g.\ Pettini 2001).  Such a model
may already encounter problems in explaining the kinematic quiescence of the
\ion{C}{4} lines (Rauch, Sargent \& Barlow 2001) but it can also be tested by
determining whether the metallicity changes with redshift.

This paper analyses the distribution of \ion{C}{4} and \ion{Si}{4} ions in the
IGM over a wide range of redshifts from $z = 1.5$\ to $z = 5.5$\ using
newly-discovered Sloan Digital Sky Survey (SDSS) quasars (Anderson et al.\
2001; Fan et al.\ 1999, 2000a,b, 2001a,b; Zheng et al.\ 2000) to provide a
high redshift sample.  I use the methodology previously applied at $z \sim 3$\
(Songaila 1997) of directly integrating the observed ion column densities in
Ly$\alpha$\ forest clouds to obtain strict lower limits on $\Omega_{\rm ion}$,
which, with plausible assumptions on ionization corrections and abundance
patterns, can provide a strict lower limit on $\Omega_{\rm metals}$\ at a
number of redshifts.

New observations of $z = 2 - 3$\ quasars with the HIRES spectrograph (Vogt
1994) on the KeckI telescope and of $z = 4 - 5.8$\ quasars with the ESI
instrument (Sheinis et al.\ 2000) on KeckII are described in section~2.  In
section~3 I construct column density distribution functions for \ion{C}{4} and
\ion{Si}{4} from $z = 1.5$\ to $z = 5.5$\ and perform the integrations.
Conclusions are given in section~4.

\section{Observations}

The data used in the paper comprise HIRES observations of 13 quasars reported
in Songaila (1998) together with HIRES and ESI observations of a further 19
quasars. The reduction of the HIRES (lower redshift) observations follows the
procedures given in Songaila (1998).  The fainter, high redshift, quasars were
observed with the ESI instrument in echellette mode. (See Songaila (2001) for
a list of quasars observed with ESI.)  The resolution in this configuration is
comparatively low, $\sim 5300$\ for the $0.75^{\prime\prime}$\ slit width
used, but the red sensitivity is high and the wavelength coverage is complete
from $4000~{\rm \AA}$\ to $10,000~{\rm \AA}$.  At the red wavelengths,
precision sky subtraction is required since the sky lines can be more than two
orders of magnitude brighter than the quasars. In order to deal with this
issue, individual half-hour exposures were made, stepped along the slit, and
the median was used to provide the primary sky subtraction. The frames were
then registered, filtered to remove cosmic rays and artifacts, and then
added. At this point a second linear fit to the slit profile in the vicinity
of the quasar was used to remove any small residual sky. The observations were
made at the parallactic angle and flux calibrated using observations of white
dwarf standards scattered through the night. These standards were also used to
remove telluric absorption features in the spectra, including the various
atmospheric bands. The final extractions of the spectra were made using a
weighting provided by the profile of the white dwarf standards. Wavelength
calibrations were obtained using third-order fits to CuAr and Xe lamp
observations at the begining and end of each night, and the absolute
wavelength scale was then obtained by registering these wavelength solutions
to the night sky lines. The wavelengths and redshifts are given in the vacuum
heliocentric frame.

I next constructed the sample of all \ion{C}{4} and \ion{Si}{4} aborption
lines which lie more than $50~{\rm \AA}$\ longward of each quasar's Lyman
alpha emission and hence clearly outside the Lyman alpha forest region.  The
doublets were found by inspection of the spectra, and confirmed by consistency
of the column density and velocity structure in the two members of the
doublet.  Systems within $5000~{\rm km\ s}^{-1}$\ of the quasar's emission
redshift were excluded to avoid contaminating the sample with proximate
systems whose ionization may be dominated by the quasar itself.  The final
\ion{C}{4} sample consists of 367 {\ion{C}{4}} doublets, with $1.74 \le z \le
5.29$\ and $12.0 \le \log (N_{\rm C~IV}) \le 14.9$.  The \ion{Si}{4} sample
contains 109 systems between $z = 1.78$\ and $z = 5.29$, with $12.0 \le \log
(N_{\rm Si~IV}) \le 14.8$.

%
%
\begin{inlinefigure}
\psfig{figure=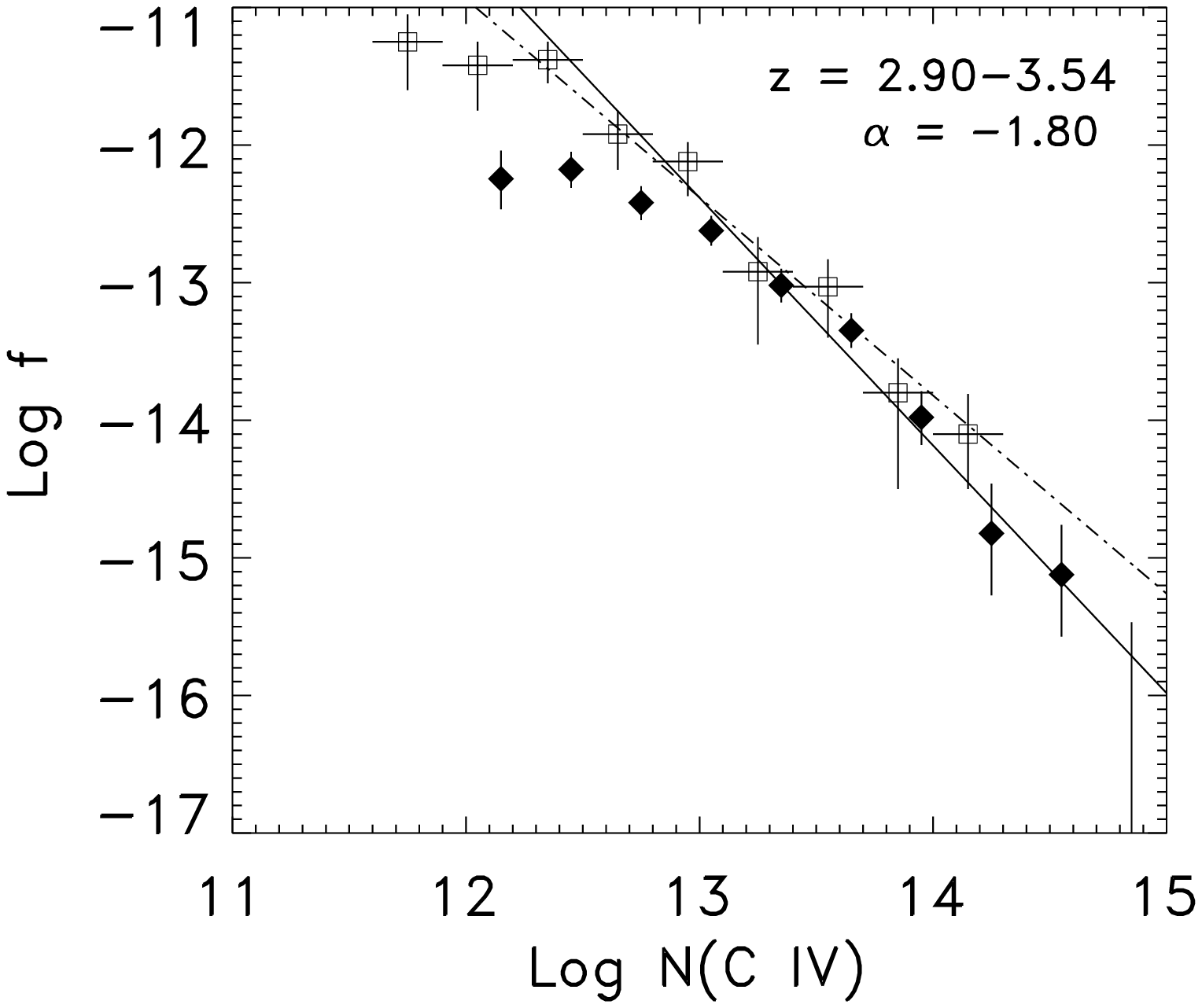,angle=90,width=4.0in}
\vspace{6pt}
\figurenum{1}
\caption{ 
Column density distribution of C~IV absorbers for the present
sample(filled symbols) in the redshift interval $2.90 < z < 3.54$; $f$ is the
number of systems per column density interval per unit redshift path.  The bin
size is $10^{0.3} N~{\rm cm}^{-2}$\ and the error bars are $\pm 1 \sigma$\
based on the number of points in each bin.  The solid line is a power law of
the form $f(N) dN = B N^{- \alpha} dN$\ with index $\alpha = 1.8 \pm 0.1$
established by a maximum likelihood fit to all systems with $N({\rm CIV}) \ge
10^{13}~{\rm cm}^{-2}$.  The open symbols are the column density distribution
of Ellison et al.\ (2000) for C~IV absorbers in the quasar Q1422+231 over the
same redshift interval (not corrected for incompleteness at low $N$).  The
dot-dash line is Ellison et al.`s power law fit with index $1.44 \pm 0.05$\ to
$N({\rm CIV}) = 10^{12}~{\rm cm}^{-2}$.  
}
\label{fig1}
\addtolength{\baselineskip}{10pt}
\end{inlinefigure}

%
%
\begin{inlinefigure}
\psfig{figure=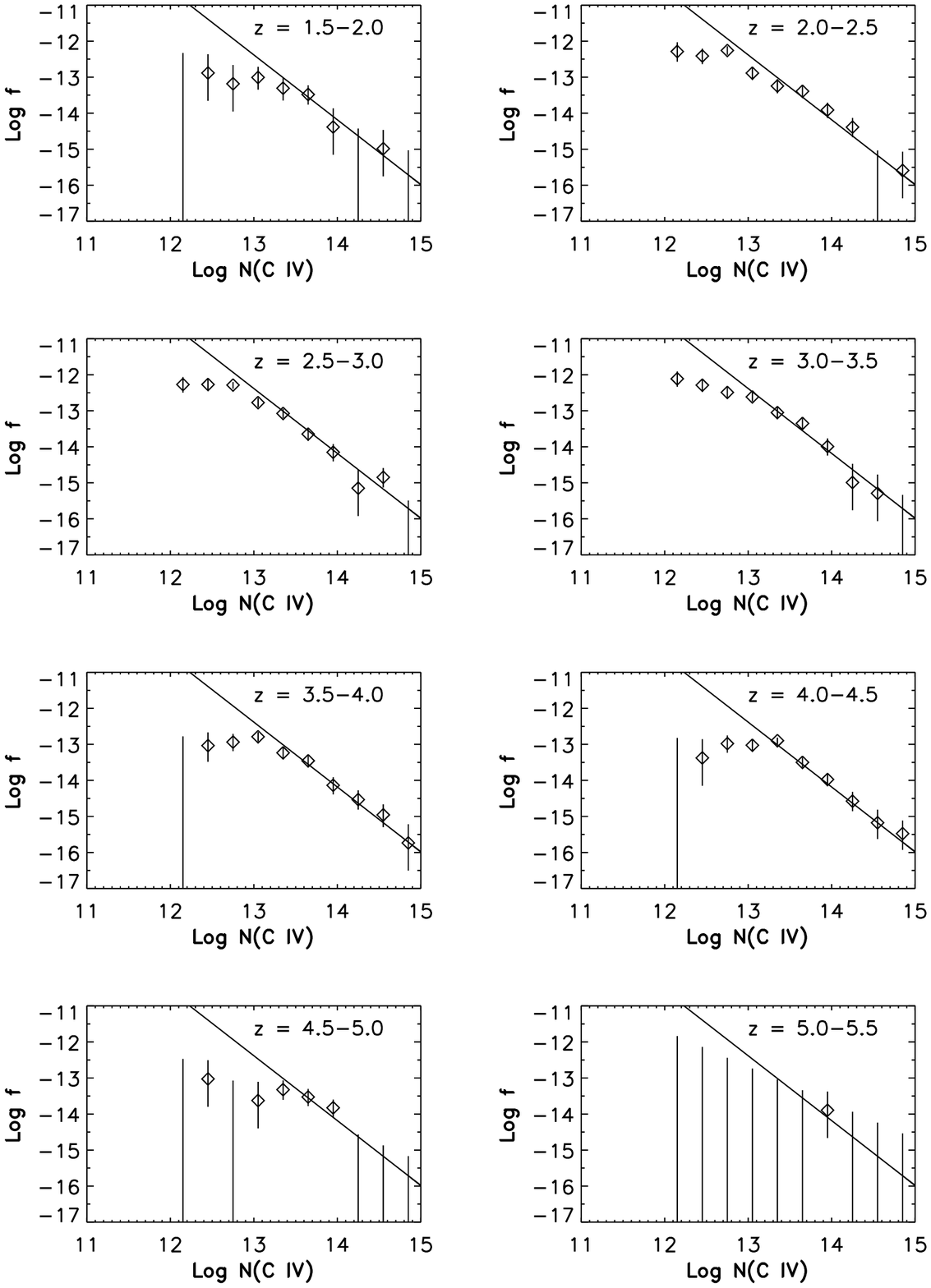,angle=0,width=3.5in}
\vspace{6pt}
\figurenum{2}
\caption{
\label{fig:2} Column density distribution of C~IV absorbers as
in Figure~1 for various redshift ranges.  In each redshift interval, the solid
line is the fiducial power law $f(N) dN = B N^{- \alpha} dN$\ with the
normalization and index $\alpha = 1.8$\ established from Figure~1.
}
\label{fig2}
\addtolength{\baselineskip}{10pt}
\end{inlinefigure}

In all cases, ion column densities were determined by fitting up to ten Voigt
profiles to each redshift system, defined to be all absorption near a given
fiducial redshift with gaps in velocity space of no more than $50~{\rm km\
s}^{-1}$.  The column density at a given redshift is then the total column
density of all such components.  In all but a few systems, individual lines
were unsaturated, so that fitted column densities are insensitive to
$b$-value. For the ESI spectra, where the lower resolution observations make
this issue more critical, I adopted in doubtful cases a column density based
on assuming that the weaker member of the doublet was unsaturated.  This
provides a minimum estimate of the column density.

\section{Distribution Functions}

With these samples, the column density distribution function $f(N)$\ was
determined for each ion.  $f(N)$\ is defined as the number of absorbing
systems per unit redshift path per unit column density, where at a given
redshift $z$\ the redshift path, $X(z)$\ is defined as 
$X(z) \equiv \case{2}{3}\,
[(1+z)^{3/2} - 1]$\ for $q_0 = 0.5$ (Tytler 1987).  Throughout the paper the
distributions are calculated in c.g.s.\ units and are plotted
with $1~\sigma$\ error bars calculated from the Poisson errors based on the
number of systems in each bin (Gehrels 1986).  In addition, all calculations
were made assuming a $q_0 = 0.5$, $\Lambda = 0$\ cosmology, which is a good
approximation to the currently favored $\Lambda$-dominated cosmologies above
about $z = 1$.  

In Figure~1 I show the column density distribution in the redshift interval
$2.90 < z < 3.54$ and compare this with the distribution over the same
redshift range from an extremely deep exposure of the quasar 1422+231, as
reported in Ellison et al.\ (2000).  From the figure it can be seen that the
present data become incomplete below a column density of approximately
$10^{13}~{\rm cm}^{-2}$\ in this redshift range.  Above this value a maximum
likelihood fit to a power law slope gives an index of $\alpha = 1.8 \pm 0.1$.
This is steeper than the Ellison et al.\ fit of $1.44 \pm 0.05$ based on the
more limited sample, but in fact provides a good fit to those data points at
column densities less than $10^{13}~{\rm cm}^{-2}$\ and to the corrected
counts in 1422+231 at less than $10^{12}~{\rm cm}^{-2}$.  I adopt this fit,
with a normalization of $\log f = -12.4$\ at a column density of $10^{13}~{\rm
cm}^{-2}$, as a fiducial to compare with the distributions in other redshift
ranges.

The {\ion{C}{4}} and \ion{Si}{4} distribution functions are shown as a function of
redshift in Figure~2.  For redshifts less than 4 the data is drawn from HIRES
observations of relatively bright quasars, and significant incompleteness
occurs only below column densities of about $10^{13}~{\rm cm}^{-2}$. At higher
redshifts, where most of the observations are from ESI and the quasars are
fainter, simulations show there is clearly incompleteness even above this
column density. In the highest redshift bin, $5 - 5.5$, lying at the
longest wavelengths where the night sky is highest, recovery of lines is quite
sporadic and the sample is extremely incomplete.  Over the column density
range in which incompleteness is not a problem, the distributions are
consistent with being invariant throughout the range of redshifts sampled.

%
\begin{inlinefigure}
\psfig{figure=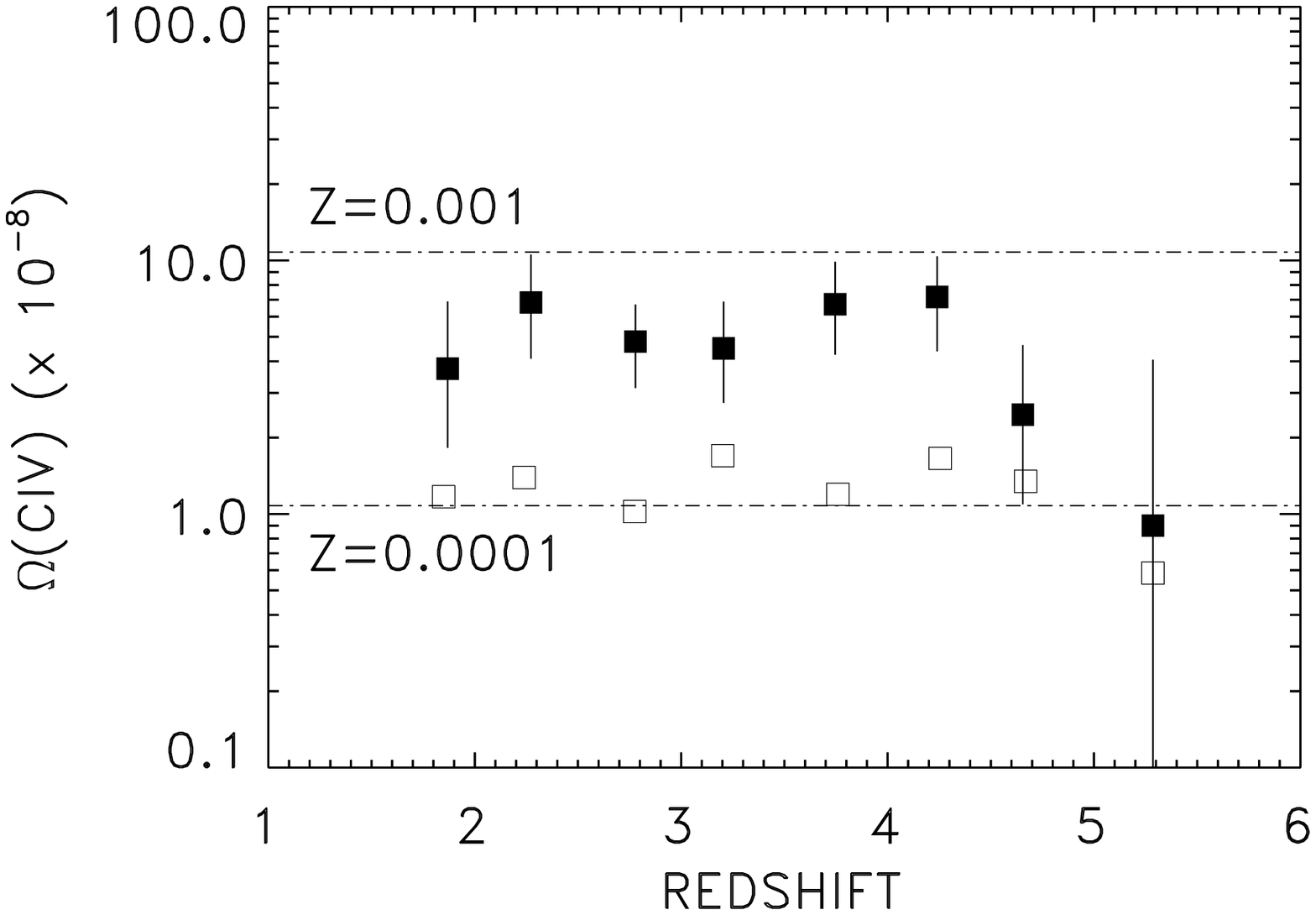,angle=90,width=3.5in}
\vspace{6pt}
\figurenum{3}
\caption{ $\Omega ({\rm CIV})$\ as a function of redshift plotted from the
data of Table~1 (${\rm H}_0 = 65~{\rm km\ s}^{-1}~{\rm Mpc}^{-1}$\ and $q_0 =
0.5$). In each redshift interval, the filled symbols show $\Omega ({\rm
CIV})$\ computed for all systems with $10^{12}~{\rm cm}^{-2} \le N({\rm CIV})
\le 10^{15}~{\rm cm}^{-2}$, whereas the open symbols are $\Omega ({\rm CIV})$
computed for all systems with $10^{13}~{\rm cm}^{-2} \le N({\rm CIV}) \le
10^{14}~{\rm cm}^{-2}$.  Symbols are plotted at the average redshift for each
bin (Table~1).  Error bars are 90\% confidence limits computed using Monte
Carlo simulations.  The dot-dash lines are $\Omega ({\rm CIV})$\ computed
assuming $\Omega_b\, h^2 = 0.023, h=0.65$ , a C~IV ionization fraction of 0.5
and metallicities $Z = 0.0001$\ and 0.001, respectively.  }
\label{fig3}
\addtolength{\baselineskip}{10pt}
\end{inlinefigure}
%
%
\begin{inlinefigure}
\psfig{figure=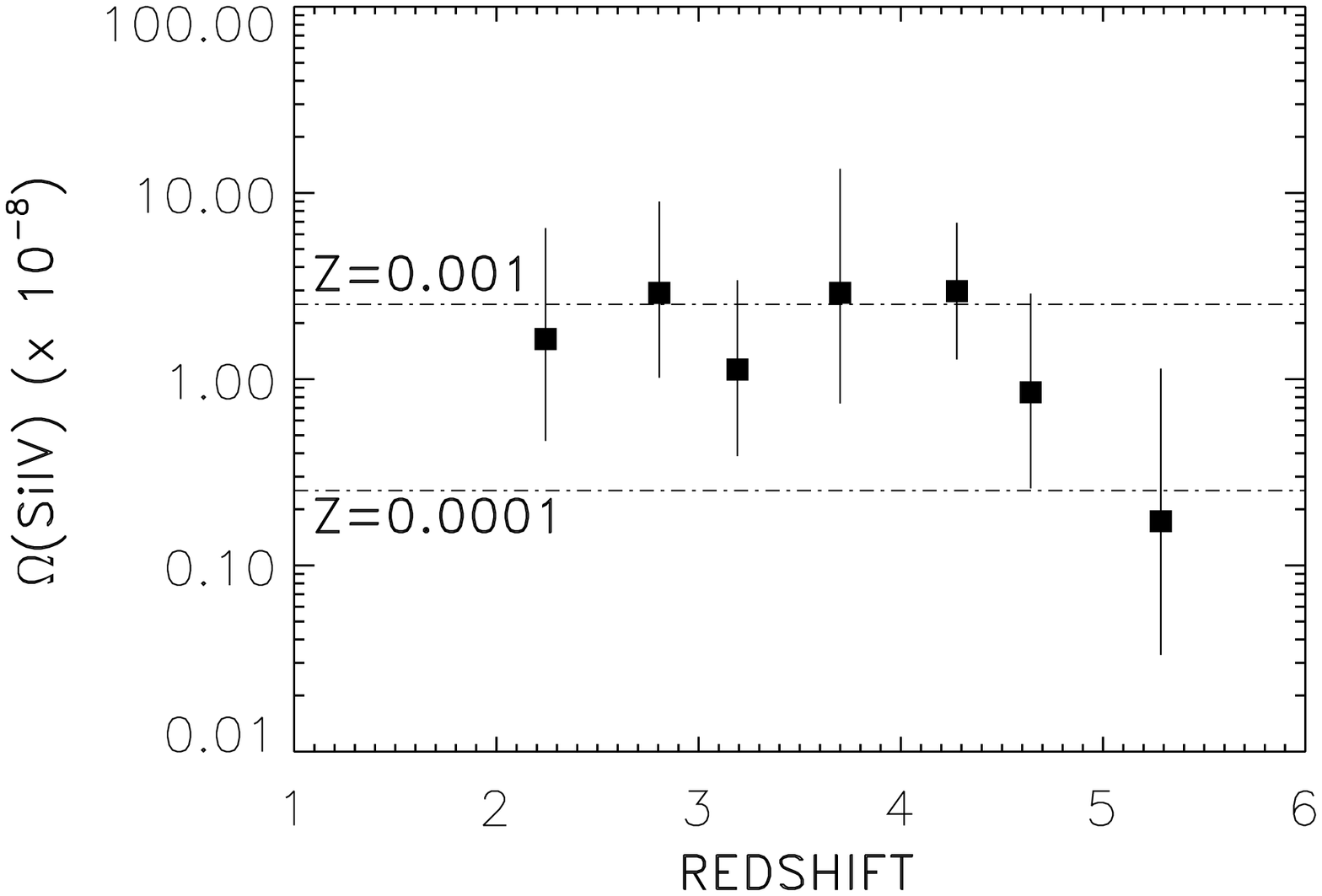,angle=90,width=3.5in}
\vspace{6pt}
\figurenum{4}
\caption{
As in Fig.~3, for Si~IV.
}
\label{fig4}
\addtolength{\baselineskip}{10pt}
\end{inlinefigure}

$\Omega_{\rm C~IV}$\ and $\Omega_{\rm Si~IV}$\ were next calculated from the 
line samples in each redshifts interval according to
\begin{equation}
\Omega_{\rm ion} = {{{\rm H}_0} \over {c\,\rho_{\rm crit}}} \  
                   \Omega_m^{1/2} \  
                   {{\sum N_{\rm ion}} \over {\Delta X}} \  
                   m_{\rm ion}
\end{equation}
\medskip\noindent
where $\rho_{\rm crit} = 1.89 \times 10^{-29} h^2~{\rm g\ cm}^{-3}$\ is the
cosmological closure density, $m_{\rm ion}$\ is the ion's mass, ${\rm H}_0 =
100h~{\rm km\ s}^{-1}\ {\rm Mpc}^{-1}$, $\Omega_m$\ is the local matter
density relative to the closure density, $dX = (1+z)^{1/2} dz$\ is the
value for a flat matter-dominated cosmology and $\Delta X$\ is the integral of
$X$\ over the redshift interval.  For $\Lambda$-dominated
cosmologies, the values may be simply scaled to preferred values of
$\Omega_m$\ using the dependence in equation~(1).  The values of $\Omega_{\rm
C~IV}$\ and $\Omega_{\rm Si~IV}$\ are plotted in Figures~3 and 4 and tabulated
in Table~1 for ${\rm H}_0 = 65~{\rm km\ s}^{-1}~{\rm Mpc}^{-1}$\ and $q_0 =
0.5$.  The error
bars are 90\% confidence limits; they were computed by using the power-law
distribution with the shape of the fiducial fit, and normalized to match the
measured $\Omega_{\rm ion}$, to generate Monte Carlo simulations of line lists
and then measuring $\Omega_{\rm ion}$ from these simulations.

\section{Discussion}

If the ionization fraction can be estimated, then $\Omega_{\rm ion}$\ can be
translated to $\Omega_{\rm element}$\ for each element in question.  I shall
assume here that \ion{C}{4} and \ion{Si}{4} are the dominant ionization stages
with ionization fractions of 0.5 (Songaila 1997). This assumption is
conservative in that it will, if anything, underestimate the resulting
metallicity since neither \ion{C}{4} nor \ion{Si}{4} can be a much larger
fraction than this; even in the extreme case, the metallicity will be
overestimated by no more than a factor of two.  The measurements of
$\Omega_{\rm element}$\ can then be compared with measurements of $\Omega_b$\
to obtain the universal metallicity contribution of the gas. (The inverse
procedure of guessing a metallicity and then using this to estimate
$\Omega_b$\ has been followed by Tripp, Savage \& Jenkins (2000) in analyzing
\ion{O}{6} measurements at low redshift.  However, given that we know
$\Omega_b$\ from other sources, it seems more reasonable in the present case
to use that value to constrain the metallicity.) A value of $\Omega_b\,h^2 =
0.023 \pm 0.003$\ is adopted from measurements of the microwave background
(Netterfield et al.\ 2001), which is consistent with the most recent deuterium
estimates (e.g., Tytler et al.\ 2001).  The metallicity is quoted relative to
solar values of $3.3 \times 10^{-4}$\ for carbon and $3.3 \times 10^{-5}$\ for
silicon (Anders \& Grevesse 1989).  The results of this procedure are
illustrated with the dashed lines in Figures~3 and 4. For $h=0.65$, I conclude
that the carbon metallicity is $5 \times 10^{-4}$\ at lower redshifts and
still exceeds $2 \times 10^{-4}$\ at $z = 5$. The silicon metallicity is about
2 times higher, $3.5 \times 10^{-4}$\ at $z = 5$, with the assumed \ion{Si}{4}
ionization fraction of 0.5.  This presumably reflects the faster generation of
the alpha process elements, but the exact amount of the excess of silicon over
carbon will of course depend on the relative \ion{Si}{4} and \ion{C}{4}
ionization fractions.

The minimum nature of the estimates should be emphazised: in any given region,
only observed metals are counted.  While we believe that the bulk of
the baryons are to be found in these clouds at these redshifts (Rauch et
al. 1997; Kim et al. 1997), the bulk of the metals may not, and may instead
be primarily retained in the galaxies that generated them.  Furthermore, the
assumption that \ion{C}{4} and \ion{Si}{4} are dominant ionization stages
depends critically on the shape and intensity of the metagalactic ionizing
radiation field and on the densities of the clouds. There may be cloud column
densities and epochs for which this assumption is no longer true and where the
observed ion densities would imply higher metallicities.  Finally, the high
noise levels at the longest wavelengths mean we are substantially
undercounting the lines at the highest redshifts, and the turndown in the ion
densities at these redshifts may be due to this effect rather than
to a real physical change.  Simulations of the incompleteness above $z = 4.5$
suggest that \ion{C}{4} line identification is substantially complete to $\log
N = 13.25$\ at $z = 4.5$\ and to $\log N = 13.7$\ above $z = 5$, resulting in
corrections to $\Omega({\rm C~IV})$\ of factors of 1.35 and 1.67, respectively, in
these two bins.

Despite these caveats we can still draw some substantial conclusions. First,
since total metal densities are dominated by oxygen, which we expect to scale
most closely with silicon, we would conclude that a minimum univeral
metallicity of about $3.5 \times 10^{-4}$\  of the solar metallicity is in
place at $z=5$. This means that this fraction of the universal massive star
formation should have taken place before this time, and this in turn can be
translated to a number of ionizing photons per baryon.  Since a metallicity of
$10^{-5}$\ gives rise to one ionizing photon per baryon (e.g., Miralda-Escud\'e
\& Rees 1997), the metallicity measured at $z=5$ is sufficient to have
preionized the IGM.

Second, there appears to have been little change in the univeral ion densities
since $z=5$.  It is possible that this could be caused by matched variations
in the ionization parameter and the metallicity.  However, at least above $z
\sim 3$, where the metagalactic ionizing spectrum is likely to be
galaxy-dominated (e.g. Steidel et al.\ 2001), \ion{C}{4} and \ion{Si}{4} will
be the dominant ionization stages, and it appears more likely that the metal
densities have remained relatively constant during this period.  Although this
does not rule out the possibility of additional contributions from mechanisms
such as galactic winds, it does suggest that, to a large extent, the metals
observed in the intergalactic medium at $z=3$ have an earlier origin.

\acknowledgments This research was supported by the National Science
Foundation under grants AST 96-17216 and AST 00-98480.

\begin{deluxetable}{lcccccccc}
\tablewidth{500pt}
\tablecaption{$\Omega_{\rm CIV}$\ and $\Omega_{\rm SiIV}$\ by Redshift \label{tbl:2}}
\tablehead{
\colhead{$z$} 
& \colhead{$\Delta X$} 
& \colhead{$\langle z \rangle$}  
& \colhead{\# Lines} 
& \colhead{$\Omega_{\rm CIV}$}
& \colhead{$\Delta X$} 
& \colhead{$\langle z \rangle$}  
& \colhead{\# Lines} 
& \colhead{$\Omega_{\rm SiIV}$}
\\ [0.5ex]  
& C~IV & C~IV & C~IV
 & ($ \times 10^{-8}$) 
& Si~IV & Si~IV & Si~IV & ($ \times 10^{-8}$) 
}
\startdata
1.5 -- 2.0 & 3.54  & 1.869 & 14 & 3.74 & \nl
2.0 -- 2.5 & 7.22  & 2.272 & 59 & 6.83 & 3.67  & 2.243 & 13 & 1.65 \nl
2.5 -- 3.0 & 11.08 & 2.779 & 88 & 4.79 & 6.05  & 2.806 & 15 & 2.90 \nl
3.0 -- 3.5 & 7.46  & 3.206 & 67 & 4.50 & 5.58  & 3.192 & 28 & 1.13 \nl
3.5 -- 4.0 & 10.49 & 3.746 & 55 & 6.71 & 2.66  & 3.700 &  8 & 2.91 \nl
4.0 -- 4.5 & 11.93 & 4.240 & 67 & 7.18 & 9.69  & 4.277 & 31 & 2.97 \nl
4.5 -- 5.0 & 5.36  & 4.655 & 16 & 2.46 & 4.35  & 4.642 &  9 & 0.85  \nl
5.0 -- 5.5 & 1.25  & 5.285 & 1  & 0.90 & 1.25  & 5.285 &  1 & 0.17


\enddata

\end{deluxetable}


\begin{references}

\reference{} Abel, T., Anninos, P, Norman, M. L. \& Zhang, Y. 1998, ApJ, 508,
518. 

\reference{} Anders, E., \& Grevesse, N. 1989, Geochim. Cosmochim. Acta, 
  53, 197.

\reference{} Anderson, S. F., et al.\ 2001, AJ, 122, 503.

\reference{} Becker, R. H., et al.\ 2001, preprint (astro-ph/0108097).

\reference{} Burles, S. \& Tytler, D. 1996, 460, 584.

\reference{} Carr, B. J., Bond, J. R. \& Arnett, W. D. 1984, ApJ, 277, 445.

\reference{} Cowie, L. L., Songaila, A., Kim, T.-S., \& Hu, E. M. 1995,
\aj, 109, 1522.

\reference{} Djorgovski, S. G., Castro, S. M., Stern, D. \& Mahabel,
A. A. 2001, preprint (astro-ph/0108069).

\reference{} Ellison, S. L., Songaila, A., Schaye, J. \& Pettini, M. 2000, AJ,
120, 1175.

\reference{} Fan, X., et al.\ 1999, AJ, 118, 1.

\reference{} Fan, X., et al.\ 2000a, AJ, 119, 1.

\reference{} Fan, X., et al.\ 2000b, AJ, 120, 1167.

\reference{} Fan, X., et al.\ 2001a, AJ, 121, 31.

\reference{} Fan, X., et al.\ 2001b, AJ, 121, 54.

\reference{} Fan, X. et al.\ 2001c, preprint (astro-ph/0108063).

\reference{} Gehrels, N. 1986, ApJ, 303, 336.

\reference{} Gnedin, N. Y., \& Ostriker, J. P. 1997, \apj, 486, 581.

\reference{} Gnedin, N. Y. 1998, MNRAS, 294, 407.

\reference{} Haiman, Z. \& Loeb, A. 1997, ApJ, 483, 21.

\reference{} Kim, T.-S., Hu, E. M., Cowie, L. L. \& Songaila, A. 1997,
AJ, 114, 1.

\reference{} Madau, P., Ferrara, A. \& Rees, M. J. 2001, ApJ, 555, 92.

\reference{} Miralda-Escud\'e, J. \& Rees, M. J. 1997, ApJ 478, L57.

\reference{} Netterfield, C. B., et al.\ 2001, preprint (astro-ph/0104460). 

\reference{} Ostriker, J. P. \& Gnedin, N. Y. 1996, ApJ, 472, L63.

\reference{} Pettini, M. 2001, in Chemical Enrichment of the Intracluster
and Intergalactic Medium, ed. F. Matteucci, (San Francisco: ASP), 
in press.

\reference{} Rauch, M., Haehnelt, M. G. \& Steinmetz, M. 1997, \apj, 481,
601. 

\reference{} Rauch, M., Sargent, W. L. W. \& Barlow, T. A. 2001, ApJ, 554,
823. 

\reference{} Sheinis, A. I., Miller, J. S., Bolte, M. \& Sutin, B. M. 2000,
Proc. SPIE, 4008, 522.

\reference{} Songaila, A. 1997, \apj, 490, L1.

\reference{} Songaila, A. 1998, \aj, 115, 2184.

\reference{} Songaila, A., Hu, E. M., Cowie, L. L. \& McMahon, R. G. 1999,
ApJ, 525, L5.

\reference{} Songaila, A. 2001, in preparation.

\reference{} Steidel, C. C., Pettini, M. \& Adelberger, K. L. 2001, ApJ, 546,
665. 

\reference{} Tripp, T. M., Savage, B. D. \& Jenkins, E. B. 2000, ApJ, 534,
L1. 

\reference{} Tytler, D., Fan, X.-M., Burles, S., Cottrell, L., Davis, C., 
  Kirkman, D., \& Zuo, L. 1995, in QSO Absorption Lines, ed. G. Meylan 
  (Heidelberg: Springer), 289.

\reference{} Tytler, D. 1987, ApJ, 321, 49.

\reference{} Tytler, D., O'Meara, J. M., Suzuki, N. \& Lubin, D. 2001, Phys.
Scr., in press.

\reference{} Vogt, S. S. et al.\ 1994, Proc. SPIE, 2198, 362.

\reference{} Zheng, W., et al.\ 2000, AJ, 120, 1607.


\end{references}
\end{document}